# Lung Nodule-SSM: Self-Supervised Lung Nodule Detection and Classification in Thoracic CT Images

Muniba Noreen and Furqan Shaukat*

Faculty of Electrical and Electronics Engineering, University of Engineering and Technology, Taxila, Pakistan

furqan.shoukat@uettaxila.edu.pk

**Abstract.**

Lung cancer remains among the deadliest types of cancer in recent decades, and early lung nodule detection is crucial for improving patient outcomes. The limited availability of annotated medical imaging data remains a bottleneck in developing accurate computer-aided diagnosis (CAD) systems. Self-supervised learning can help leverage large amounts of unlabeled data to develop more robust CAD systems. With the recent advent of transformer-based architecture and their ability to generalize to unseen tasks, there has been an effort within the healthcare community to adapt them to various medical downstream tasks. Thus, we propose a novel "*LungNodule-SSM*" method, which utilizes self-supervised learning with DINOv2 as a backbone to enhance lung nodule detection and classification without annotated data. Our methodology has two stages: firstly, the DINOv2 model is pre-trained on unlabeled CT scans to learn robust feature representations, then secondly, these features are fine-tuned using transformer-based architectures for lesion-level detection and accurate lung nodule diagnosis. The proposed method has been evaluated on the challenging LUNA 16 dataset, consisting of 888 CT scans, and compared with SOTA methods. Our experimental results show the superiority of our proposed method with an accuracy of 98.37%, explaining its effectiveness in lung nodule detection. The source code, datasets, and pre-processed data can be accessed using the link: https://github.com/EMeRALDsNRPU/Lung-Nodule-SSM-Self-Supervised-Lung-Nodule-Detection-and-Classification/tree/main

## 1    Introduction

Lung cancer is the most widely diagnosed cancer and the leading cause of cancer-related deaths worldwide [1, 2].In 2024, approximately 2.5 million new cases occurred globally, accounting for 12.4% of all new cancer diagnoses [3, 4]. In the United States, lung cancer remains a significant health concern, with an expected 226,650 new cases and 124,730 deaths likely in 2024 [5]. Particularly, lung cancer incidence rates are nearly twice as high in males compared to females, with the highest rates observed in Central and Eastern Europe, as well as in Southern Europe, Western Asia, and Melanesia [6]. Early detection through low-dose CT scans has improved survival rates, though screening acceptance remains low, with only 16.0% of those eligible going through screening in 2022 [7].

## 1.1 Background

Medical imaging methods like Computed Tomography (CT) and X-rays are vital for detecting and classifying lung nodules [8]. Given the frequent use of CT scans and the intricate nature of medical images[9], Computer-Aided Detection (CADe) and Computer-Aided Diagnosis (CADx) significantly aid radiologists by improving the precision of lung nodule detection and diagnosis [10].CAD systems can help clinicians [11] analyze medical images to identify nodules. The five-year survival rate for lung cancer is recorded to be only 25% [12]. With such a low survival rate, there is a need to create an automated system that assists in the early detection of lung nodules for improved follow-up procedures. Conventional image processing methods, including thresholding, region-based approaches, and edge detection, struggle with the complexities inherent in medical imaging [13].

Deep Learning is crucial for detecting and classifying lung cancer, leading to improved treatment and follow-up procedures [14]. These models extract features from medical images and accurately segment nodules in CT scans and X-rays. Recent advancements in deep learning have significantly boosted the analysis performance in medical imaging[15]. While Convolutional Neural Networks (CNNs) excel in analyzing complex medical images, they struggle with 3D data. To address this, 3D CNN architectures [16] have emerged to effectively handle the dimensions of 3D scans, allowing for the extraction of significant features. However, extensive training is required for these models to understand the complex structures in medical images [17]. For example, DHEA-Net [18] features a dual encoder-based architecture that utilizes CT scans and coronal views for enhanced segmentation accuracy. A systematic review and meta-analysis assessed the diagnostic accuracy of deep learning models for lesion-wise sensitivity [19].

Self-supervised learning (SSL)[20] has emerged as a robust method for pretraining on large unlabeled datasets, enabling models to create rich, task-agnostic feature representations that generalize effectively to new domains. SSL methods integrating contrastive objectives with masked image pretraining have shown substantial improvements in medical image analysis, especially in detecting pulmonary nodules using the LUNA16 challenge dataset[21]. However, the data-intensive nature of these methods presents challenges for training and generalizability. Producing detailed labels to train deep learning models for 3D medical imaging is time-consuming and costly. SSL techniques have been proposed to reduce dependency on detailed ground truth annotations by utilizing unlabeled datasets [22, 23]. Until now, most existing approaches to 3D medical imaging modalities train SSL methods using simple pretext formulations on unlabeled datasets like their downstream applications. There was still a gap in guidance on selecting the most suitable pretext task for fine-tuning. So, the pretrain-then-finetune workflow is recommended for improving performance. This enables the model to pretrain with valuable properties that may be adjusted for downstream tasks. Although these models are well-developed for general segmentation and domain adaptation, lesion-level detection and nodule characterization in lung cancer diagnosis have not been explored yet.

To address this issue, we aim to enhance lung nodule detection accuracy and efficiency using a self-supervised learning method. Building on this objective, our research introduces an advanced method for identifying the lesion characteristics and features using the Dinov2 model. Our approach delivers accurate and efficient results for early diagnosis and effective treatment. Our contributions are summarized as follows:

1. Developed a fully automated, end-to-end lung nodule detection and classification pipeline using a self-supervised learning (SSL) approach.
2. Enhanced detection and classification performance on lung CT images by leveraging DINOv2-based Vision Transformer features.
3. Achieved significant improvements in parameter efficiency and classification accuracy by fine-tuning the pretrained transformer encoder.

## 2      Related Work

Research on the detection and classification of automatic lung nodules has evolved through three main approaches: volumetric convolutional neural networks (3D CNNs), transformer-based self-supervised learning (SSL) methods, and SSL strategies designed explicitly for the LUNA16 dataset. In earlier studies, Ilesanmi et al. utilized 3D U-Net and V-Net architectures to capture inter-slice dependencies in CT volumes, significantly enhancing segmentation accuracy by maintaining spatial continuity via skip connections. However, this method requires large amounts of annotated data and substantial GPU memory, which can limit its feasibility in data-scarce environments [24]. Usman et al. recently developed DHEA-Net, a dual-encoder framework that integrates axial and coronal views through a specialized fusion module to produce more refined segmentation outputs. Though DHEA-Net shows improved accuracy, its dual encoder architecture increases model parameters and resource demands, creating challenges for deployment in institutions with restricted computational capabilities [25].

The emergence of Vision Transformers (ViTs) and self-supervised objectives has driven the development of transformer-based methods in medical imaging. Wang et al. modified SimCLR for lung CT by ensuring alignment between augmented patch views and fine-tuning a DenseNet121 backbone on the resulting embeddings; however, the extensive augmentation method may introduce anatomically unrealistic artifacts and lengthen training time [26]. Zhou et al. utilized Momentum Contrast (MoCo) with a ResNet50 encoder, using a dynamic memory bank for self-supervision. At the same time, successful on unlabeled datasets, MoCo's dependence on large feature queues complicates its integration into real-time clinical environments[27]. Li and Yang's BYOL-based approach does without negative pairs by employing dual-network self-distillation. Yet, its effectiveness is significantly influenced by meticulous momentum and predictor scheduling to avoid collapse [28]. The DINOv2 framework (Oquab et al., 2023) has further progressed self-supervised learning by merging self-distillation with Masked pretraining on ViTs[29], demonstrating zero-shot nodule detection capabilities, as highlighted by Cai and Xu[30]. A pretrained DINOv2 model achieves 92.3% accuracy without needing domain-specific fine-tuning. However, it struggles to recognize the subtle radiographic textures unique to CT data.

Ultimately, the LUNA16 dataset has established itself as the standard benchmark for pulmonary nodule SSL. Liu et al. utilized a contrastive SSL pipeline on LUNA16 volumes, sampling 3D patches to develop feature embeddings that were later classified using a simple multilayer perception. While their approach improved sensitivity compared to supervised baselines, the absence of global context aggregation across slices led to a plateau in performance [21]. These studies highlight that although volumetric CNNs excel in context modeling and transformer-based SSL offers robust features, there remains a gap in integrating domain-tuned fine-tuning with lightweight classifiers. This inspires our strategy: to fine-tune DINOv2 on LUNA16 and combine its embeddings with a Random Forest classifier, thereby achieving a balance between representational power, interpretability, and efficient form in computations.

## 3   Methodology

In this methodology section, we discussed our study to enhance the efficiency of lung nodule detection and classification using self-supervised learning techniques. The process includes dataset curation, model training utilizing the DINOv2 model[31], and performance evaluation. The DINOv2 model, using a vision transformer architecture[32, 33], was initially pre-trained on a large dataset to capture visual features. Consequently, the model is fine-tuned on the LUNA16 dataset[34], a publicly available collection of CT scans for lung nodule detection and classification. This fine-tuning process enabled the model to learn the good features and patterns associated with the lung nodule region. Finally, the algorithm's performance was evaluated using a validation set of data to check its efficacy in accurately identifying and classifying the lung nodule.

### 3.1   Dataset curation and pre-processing

We have utilized the Lung Nodule Analysis 2016 (LUNA16) challenging dataset[34] for training and validation purposes, consisting of 888 scans. The nodule inclusion criteria from LUNA16 were observed for the following evaluations, which identified 1,186 nodules. In preprocessing, we divided each CT scan into smaller, uniform 2D patches to effectively leverage the rich spatial information in the scans. These 2D patches served as input for training the DINOv2 model, enabling unsupervised feature learning through self-supervision. The learned features were fine-tuned to classify lung nodules in downstream tasks accurately.

### 3.2   Network Architecture

Figure 1 shows the block diagram of our proposed model, which consists of two stages: feature extraction and fine-tuning. In the first stage, we use a pretrained DINOv2 encoder to extract features from CT images. In contrast, the second stage involves fine-tuning task-specific heads for lung nodule detection and classification.

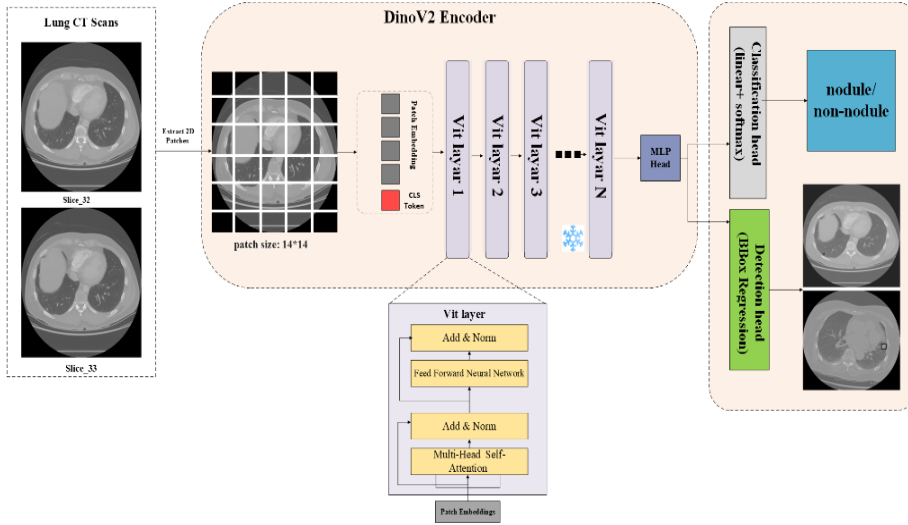

Figure 1 The proposed method includes two stages. In Stage 1, lung CT slices are processed through a DINOv2 pre-trained model to extract robust features. In Stage 2, these features are fine-tuned through task-specific heads for nodule classification and precise localization via bounding box regression.

### 3.2.1 Feature Extraction

We have used the pre-trained DINOv2 model to extract robust features. DINOv2 is a Transformer-based architecture (ViT). DINOv2 is attributed to having a unique ability for self-supervised learning and efficient training on unlabeled datasets. This model performs well in various tasks of image processing applications, such as image classification, object detection, and segmentation, with robust feature extraction capabilities. Researchers are increasingly exploring using DINOv2 for medical image analysis tasks [35].

**Patch Embeddings:** The original DINOv2 is a 1B parameter model trained on 142 million natural images with a 224 x 224 pixels resolution. We used the ViT-l/14 architecture to illustrate the detection workflow with DINOv2. It extracts the first set of features that the model learns from the images. The input CT slice consists of 512×512 pixels and is then resized to 504×504 pixels to preserve high-resolution detail to meet the patching scheme of the ViT-L/14 model. The image is then split into 1296 non-overlapping 14×14 patches, embedded into 1536-dimensional tokens, and passed through a series of transformer layers in the DINOv2 encoder. The transformer layers give high-dimensional representations of the images, which are combined to produce a feature vector. After extracting the feature vector, global average pooling is utilized to average the feature map over its spatial dimensions, resulting in the final output.

**DINOv2 Encoder:** DINOv2's Vision Transformer (ViT) Backbone is a structured approach for feature extraction, employing self-attention mechanisms to process images efficiently. The transformer encoder consists of several self-attention layers and feed-forward networks (FFN), which allow the model to understand long-range relationships and intricate spatial connections throughout the data. After passing through these layers, a classification token (CLS token) aggregates global feature representations, which are then passed into a classifier head for final prediction. [36].

### 3.2.2    Fine-tuning:

DINOv2 has proven highly effective in machine learning [37]. DINOv2 has effectively reduced the number of trainable parameters [38], enabling a wide range of applications without the high costs typically associated with traditional fine-tuning. The extracted features are fed into two heads: a classification head for nodule versus non-nodule classification and a detection head for bounding box regression.

*Classification Head:* After the encoding process, the class token is passed through a final fully connected layer, which outputs predictions for nodule or non-nodule.

*Detection Head:* To prepare inputs for the detection head, we used voxel coordinates for precise nodule localization within the CT image. A centered patch was extracted using the bounding voxel range, and the mid-axial slice was selected and replicated across all three channels to simulate 3D spatial context in a 2D format. This strategy ensured compatibility with the pretrained ViT-DINOv2 model while preserving the essential spatial characteristics of lung nodules for effective feature extraction and bounding box regression. A voxel in a 3D CT scan is represented by its spatial coordinates (x, y, z), as shown in Eq. 1. The window size used to extract a 3D region is denoted as w in Eq. 2, and I represents the final image with a bounding box in Eq. 3.

$$V(x, y, z) \in R \quad (1)$$

To extract a lung nodule centered in the CT volume:

$$v_n = V\left(x_C - \frac{w}{2} : x_C + \frac{w}{2}\right), \left(y_C - \frac{w}{2} : y_C + \frac{w}{2}\right), \left(z_C - \frac{w}{2} : z_C + \frac{w}{2}\right) \quad (2)$$

For 2D ViT processing, selecting the mid-axial slice:

$$I = V(x_C - \frac{w}{2} : x_C + \frac{w}{2}, y_C - \frac{w}{2} : y_C + \frac{w}{2}, z_c) \quad (3)$$

### 3.3  Loss Function

**Nodule Classification:** For classification as nodule or non-nodule, we used the loss function defined as:

$$L_{SCE} = \alpha \cdot L_{CE}(p, q) + \beta \cdot L_{RCE}(p, q) \quad (4)$$

Where:

$$L_{CE}(p, q) = - \sum_i p_i \log(q_i) \quad (5)$$

In Eq. 4, the cross-entropy loss is denoted, α and β are the weighting coefficients for the cross-entropy loss and the reverse cross-entropy loss, respectively. In Eq. 5, $p_i$ and $q_i$ represent the predicted and actual probabilities.

## 4    Experimental Results and Discussion

The experimental design of this study comprises two main phases. Initially, we optimized the self-supervised DINOv2 (ViT) model with the LUNA16 dataset to obtain highly distinguishing features from lung CT slices without manual annotations. These features provide valuable contextual and semantic insights for precise lung nodule identification. In the subsequent phase, we leveraged the learned feature embeddings for classification. Specifically, we extracted deep features from the optimally tuned DINOv2 model. We employed various classical classifiers, Decision Tree, Random Forest, and K-Nearest Neighbors (KNN), to differentiate between regions with nodules and non-nodules. This methodology allowed us to assess the robustness of DINOv2 embeddings across several conventional machine-learning approaches.

We established a clear experimental protocol to ensure a comprehensive and reproducible evaluation of our lung nodule detection pipeline based on DINOv2. The dataset was divided into 70% for training, 15% for validation, and 15% for testing. Each CT scan underwent preprocessing by normalizing intensity values, resizing to 504×504 pixels, and segmenting into non-overlapping 14×14 patches to comply with the input requirements of the ViT-L/14 architecture. For training, we adopted the Adam optimizer with an initial learning rate of 0.0001, a batch size of 32, and a weight decay of 0.01. The model was trained for 100 epochs on an NVIDIA RTX 3090 GPU, with an average training time of roughly 6 minutes per epoch. These parameters were chosen following empirical tuning to ensure both performance and stability.

Recent advancements in self-supervised learning (SSL) have significantly impacted lung nodule detection and classification. Table 1 illustrates the comparison with the following studies: Chen. et al using convolutional neural networks (CNNs), achieving an accuracy of 78.5%. However, their method faced challenges in capturing the complex spatial features present in 3D medical images. Zhou et al. (2022) employed Momentum Contrast (MoCo) with a ResNet50 backbone, attaining an accuracy of 85.6%. Nevertheless, the dependence on 2D CNN architectures limited the model's ability to utilize 3D spatial information fully. Wang et al. (2023) used SimCLR with DenseNet121, achieving 88.2% accuracy but requiring extensive data augmentation and computational resources. Li et al. (2023) applied BYOL with EfficientNet-B0, reaching an accuracy of 90.1%; thus, the complexity of the model presented deployment challenges in resource-constrained environments. Cai et al. (2024) adopted DINOv2 with Vision Transformers (ViT), attaining an accuracy of 92.3%; however, the lack of fine-tuning on domain-specific data limited the model's adaptability to subtle variations in lung nodule characteristics. In contrast, our proposed method utilized DINOv2 ViT architecture, fine-tuned on the LUNA16 dataset. It incorporates classifiers such as Decision Trees, Random Forests, and K-Nearest Neighbors, achieved an impressive accuracy of **98.37%** with Random Forest, as illustrated by the performance of the classifiers in Table 2.

In Figure 2, we present three representative axial CT slices, illustrates the practical identification of nodules, complete with accurate bounding boxes, despite the differing anatomical features. These examples highlight the pipeline's ability to generalize and localize nodules throughout various regions within a CT volume.

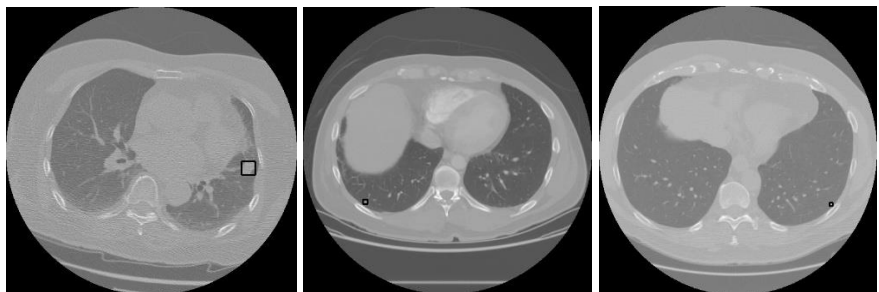

Figure 2: Few samples of results generated using purposed method

Table 1: Comparison of Lung Nodule Detection and Classification Results with Existing SSL Methods

| Author | Year | Method | No. of Samples | Accuracy (%) | F1 Score (%) | Precision (%) | Recall (%) |
|---|---|---|---|---|---|---|---|
| Chen et al. [39] | 2021 | Contrastive Learning (CNN) | 800 CT | 78.5 | 75.0 | 76.2 | 74.3 |
| Zhou et al. [40] | 2022 | MoCo (ResNet50) | 888 CT | 85.6 | 82.0 | 84.5 | 79.3 |
| Wang et al. [41] | 2023 | SimCLR (DenseNet121) | 1018 CT | 88.2 | 84.0 | 85.9 | 82.2 |
| Li et al. [42] | 2023 | BYOL (EfficientNet-B0) | 1018 CT | 90.1 | 87.0 | 88.4 | 85.6 |
| Cai et al. [43] | 2024 | DINOv2 (ViT) | 888 CT | 92.3 | 89.0 | 91.5 | 87.8 |
| **Ours** | 2025 | **DINOv2+RF** | 888 CT | **98.37** | **96.63** | **99.64** | **93.48** |

Table 2: Results on LUNA16 Dataset Using Purposed Method

| Classifier | Accuracy (%) | Precision (%) | Recall (%) | F1-score (%) | AUC (%) |
|---|---|---|---|---|---|
| Decision Tree | 81.66 | 66.21 | 54.39 | 59.72 | 80.78 |
| Random Forest | 98.37 | 99.64 | 93.48 | 96.63 | 98.09 |
| KNN | 94.76 | 99.64 | 79.32 | 88.33 | 96.87 |

## 5   Conclusions

Self-supervised learning models have demonstrated significant potential in overcoming the limitations of labeled data, offering improved generalizability and robustness, two critical aspects of medical image analysis. This study explored the effectiveness of the DINOv2 pre-training technique on the LUNA16 lung CT dataset. The model showed exceptional performance in nodule detection and classification tasks, achieving the highest accuracy of **98.37%.** DINOv2's ability to learn robust and transferable representations without supervision makes it a compelling approach for building scalable and efficient computer-aided diagnosis systems. Our results suggest that integrating self-supervised techniques, such as DINOv2, can significantly enhance the performance and reliability of deep learning models in medical imaging, paving the way toward developing large-scale medical foundation models. Future research will focus on improving the segmentation of small nodules, integrating multi-modal imaging, and optimizing algorithms to satisfy real-time clinical throughput needs. Together, these advancements will move us toward scalable tools with low annotation costs for early lung cancer detection and diagnosis.

**Acknowledgments.** This work is part of the NRPU **project# 17019** entitled "EMeRALDS: Electronic Medical Records driven Automated Lung nodule Detection and cancer risk Stratification" funded by the Higher Education Commission of Pakistan.

**Disclosure of Interests.** The authors have no competing interests to declare that are relevant to the content of this article.